\documentclass[iop,apj,twocolappendix]{emulateapj}
\usepackage{graphicx}% This allows you to call \includegraphics
\usepackage{amsmath,natbib}
\def\Omga{$\boldsymbol\Omega$}

\def\etal{{\it et al}}

\def\P3hat{{\mathaccent 94 P}_3}

\def\eg{{\it e.g.}}

\def\aap{A\&A}

\def\apjs{{\it Ap.J. Suppl.}}

\newcount\notenumber
\def\clearnotenumber{\notenumber=0}
\def\note{\advance\notenumber by1 \footnote{$^{\the\notenumber}$}}
\clearnotenumber
\usepackage{rotating}   % to rotate the expressions and tables.

\begin{document}

\title{TOWARD AN EMPIRICAL THEORY OF PULSAR EMISSION XI. UNDERSTANDING THE ORIENTATIONS OF PULSAR RADIATION AND SUPERNOVA ``KICKS''}

\author{Joanna M. Rankin}
\affil{Physics Department, University of Vermont, Burlington, VT 05405 USA: \\Joanna.Rankin@uvm.edu}

\shorttitle{Orientations of Pulsar Radiation and Supernova ``Kicks''}
\shortauthors{Joanna M. Rankin}

\begin{abstract}
Two entwined problems have remained unresolved since pulsars were discovered  
nearly 50 years ago:  the orientation of their polarized emission relative to the emitting 
magnetic field and the direction of putative supernova ``kicks' relative to 
their rotation axes.  The rotational orientation of most 
pulsars can be inferred only from the (``fiducial') polarization angle of their radiation,  
when their beam points directly at the Earth and the emitting polar fluxtube field is 
$\parallel$ to the rotation axis.  Earlier studies have been unrevealing owing to the 
admixture of different types of radiation (core and conal, two polarization modes), 
producing both $\parallel$ or $\perp$ alignments.  In this paper 
we analyze the some 50 pulsars having three characteristics:  core radiation beams, 
reliable absolute polarimetry, and accurate proper motions.  The ``fiducial' polarization 
angle of the core emission, we then find, is usually oriented $\perp$ to the proper-motion 
direction on the sky.  As the primary core emission is polarized $\perp$ to the projected 
magnetic field in Vela and other pulsars where X-ray imaging reveals the orientation, 
this shows that the proper motions usually lie $\parallel$ to the rotation axes on the sky.   
Two key physical consequences then follow:  first, to the extent that supernova ``kicks'  
are responsible for pulsar proper motions, they are mostly $\parallel$ 
to the rotation axis; and second that most pulsar radiation is heavily processed 
by the magnetospheric plasma such that the lowest altitude ``parent' core emission is 
polarized $\perp$ to the emitting field, propagating as the extraordinary (X) mode.

\end{abstract}


\section{Introduction}
\begin{figure}
\begin{center}
\includegraphics[width=86mm]{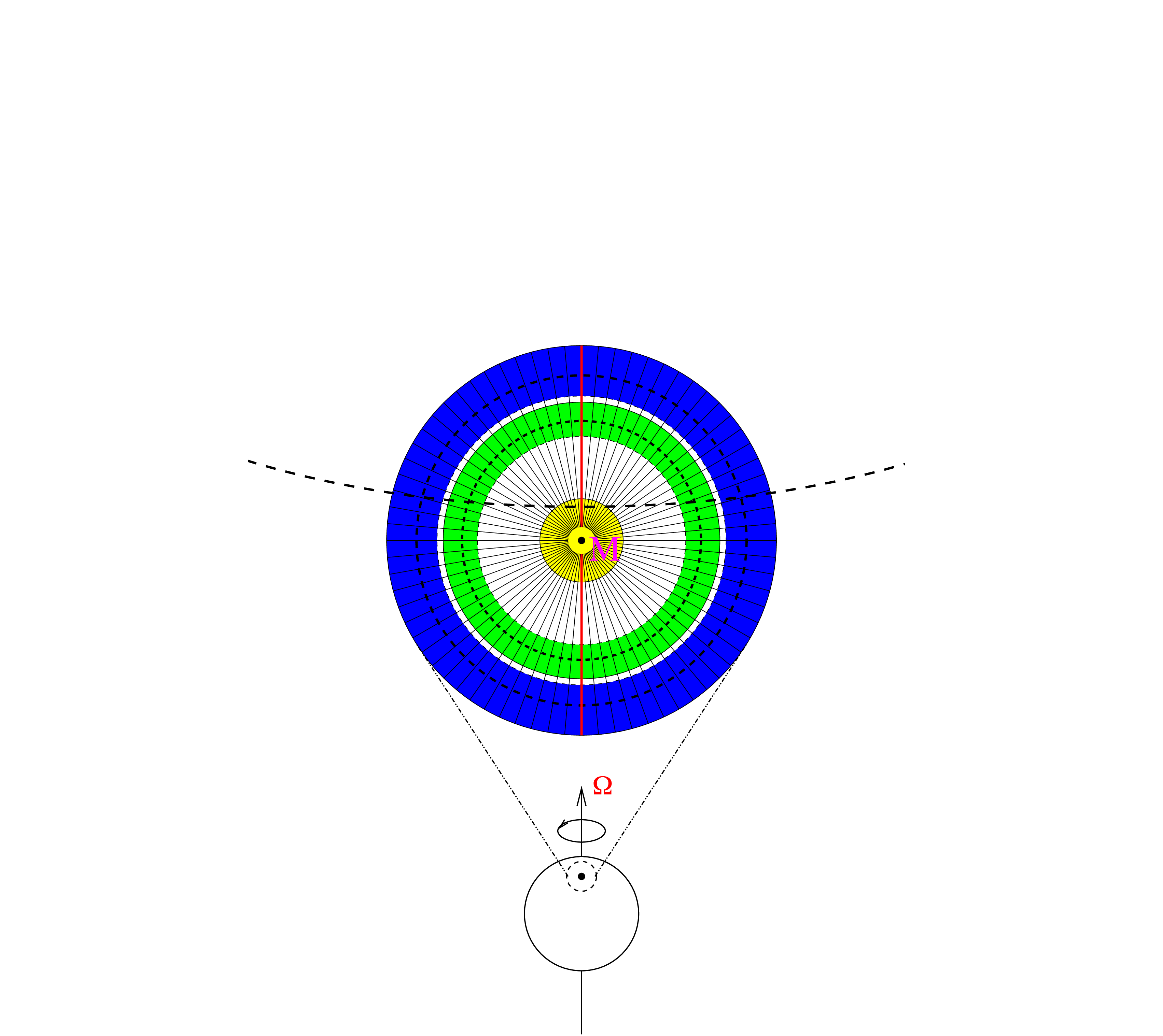}
\caption{Diagram showing the splayed curving field lines---here shown in 
projection---associated with pulsar emission at the ``fiducial'' instant when 
the beam faces the Earth directly.  Both the magnetic axis {\bf M} and pulsar 
rotation axis \Omga\ are indicated in relation to the ``fiducial''  field lines 
(magneta) associated with the radiation we receive at that instant.  The 
projected fiducial field is $\parallel$ to the rotation axis \Omga\ and thus 
used as a proxy for \Omga\ which is otherwise usually not observable.  
Also indicated is the emission-beam structure consisting of two concentric 
conal beams (outer: blue and inner: green) and the central core beam (yellow) 
in relation to a typical sightline traverse (black dashed).  Any of the three beams 
can be present in a given pulsar and different sightline paths result in different 
profile types.  Profiles with a core beam are of primary interest here---that is, 
the core-single ({\bf S$_t$}) where it appears on its own; triple ({\bf T}) profiles 
where one of the cones is present; and five-component ({\bf M}) profiles where 
both cones appear.}
\label{fig1}
\end{center}
\end{figure}

Radio pulsars now contribute importantly to many fields of physical science, but paradoxically, 
two fundamental coupled issues have remained unresolved since they were discovered 
47 years ago:  the orientation of their linearly polarized emission relative to the magnetic field 
in their polar fluxtube emitting regions and the origin/orientation of their often large space 
velocities (and thus proper motions) relative to their rotation axes.  Most pulsars are known only 
by their lighthouse-like radio signals, and thus we have no direct means of determining the 
orientation of a pulsar's rotation axis on the sky, which is crucial both to interpreting the 
polarization direction of the radiation and the orientation of the proper motion relative to 
the spin axis.  Pulsars radiate because highly energetic outward-going charges are 
accelerated by the curved dipolar field within their polar regions, so it is crucial to 
understand how this radiation is polarized relative to the (projected) {\bf B} field orientation 
on the sky.  Figure~\ref{fig1} shows how this field appears splayed when a pulsar's beam 
points directly at the Earth---the ``fiducial'' instant---and that the emission reaching us is 
associated with that bundle in the plane of the rotation axis \Omga.  Clearly, we have no 
knowledge of the radial component of a pulsar's space velocity and only weak estimates 
of the radial component of \Omga.  

At one time it seemed obvious that pulsar radiation must be polarized $\parallel$ to the 
projected magnetic field direction.  How could it be otherwise given that both the curving 
{\bf B} field and the resulting curvature acceleration lie in the same plane?  Even after the 
discovery that pulsars emit in two orthogonal (hereafter OPM) polarization modes 
[\cite{1980ApJS...42..143B}; \cite{1975ApJ...196...83M}] many assumed in the absence 
of any direct proof that the ``primary'' polarization mode must be $\parallel$.  

This easy presumption was dashed in the new millennium by X-ray imaging of the Vela 
pulsar [\cite{2001ApJ...556..380H}; \cite{2001A&A...379..551R}] where arcs indicated the 
orientation of the star's rotation axis \Omga\ relative to its polarization and proper-motion 
(PM) directions.  Shockingly, the radiation was polarized $\perp$ to the projected magnetic 
field {\bf B} plane, a circumstance then beautifully confirmed for the radio emission 
\citep[hereafter Johnston I]{2005MNRAS.364.1397J}---and this pulsar's radio emission 
is almost completely linearly polarized, so there was no OPM ambiguity.  

In an earlier paper \citep[hereafter Paper I]{2007ApJ...664..443R} we investigated the PPA vs. PM 
alignments of a number of pulsars, drawing strongly on Johnston I as well as other sources.  
Here, a ``fiducial'' polarization position angle (PPA) $PA_0$, at a (``fiducial'') rotational phase 
representing the magnetic axis longitude, is measured and referred to infinite frequency 
as a proxy for the (unseen) orientation of the rotation axis \Omga.  These were compared 
with well determined proper-motion (PM) directions $PA_V$, and the differences $\Psi$ 
showed strong peaks at both 0\degr\ and 90\degr.  Given, however, that most of the pulsars 
showed strong OPM activity in their profiles, it was not possible to draw general conclusions 
about the polarization orientation with respect to the projected {\bf B} direction.  

The second coupled key question is how a pulsar's rotation vector \Omga\ is oriented 
with respect to its space velocity (of which we can usually measure only their projections 
on the plane of the sky)?  The possibility of a correlation was raised early [\cite{1969AZh....46..715S}; 
\cite{1970ApJ...160L..91G}; \cite{1975Natur.254..676T}] as their large peculiar 
velocities were realized, initially through scintillation studies.  Though no binary pulsar 
was then known, their presumptive birth in the disruption of such systems appeared to be 
a significant factor in their PMs.  However, binary disruption ultimately seemed inadequate 
to account for the very large velocities of some pulsars.  Theorists then began to explore 
other mechanisms for their acceleration---in particular the question of whether natal 
supernovae imparted ``kicks'' to their progeny.  

This ``kick'' question has a complex history which is nicely summarized in \citep{2012MNRAS.423.2736N}.   Suffice it to say that theorists suggested ingenious ``kick'' mechanisms---both 
$\parallel$ and $\perp$ to the orientation of a pulsar's rotation \Omga\ (as well possibly in 
combination?).  Mechanisms such as Galactic acceleration are known which would tend 
to alter such ``birth'' alignments \citep[\eg,][]{2013MNRAS.430.2281N}, however the analyses 
quoted above indicate that a surprisingly large proportion of pulsars show $\Psi$ orientation 
close to either 0\degr\ or 90\degr.   Again, however, the orientation of \Omga\ cannot 
usually be determined directly, but only though the fiducial PPA proxy $PA_0$ that is 
complicated by OPM radiation either $\parallel$ or $\perp$ to the projected {\bf B} direction.  
This double ambiguity of proper-motion direction and OPM orientation  has plagued efforts 
to settle how supernovae contribute to pulsar velocities.  

Although much of our earlier work had focused on classifying the properties of different 
pulsar profile populations according primarily to their emission geometry and frequency 
evolution [Rankin \citeyear{1983ApJ...274..333R}, \citeyear{1990ApJ...352..247R}, 
\citeyear{1993ApJ...405..285R}, \citeyear{1993ApJS...85..145R}; \cite{1989ApJ...346..869R}; 
\cite{2011ApJ...727...92M}; hereafter ET I, IV, VIa,b, --- \& IX], our analysis in Paper I failed 
to use this information, despite the fact that most of the stars there under consideration 
had long been classified through detailed study.  Most of the pulsars in the Johnston I 
study exhibited core, core-cone, or core-double cone profiles of the core-single ({\bf S$_t$}), 
triple ({\bf T}) or five-component ({\bf M}) types (see Fig.~\ref{fig1}), and most of the 
90\degr\ alignments pertained to pulsars of these types, but this circumstance was not 
then observed or interpreted.  

\begin{figure}
\begin{center}
\includegraphics[width=76mm]{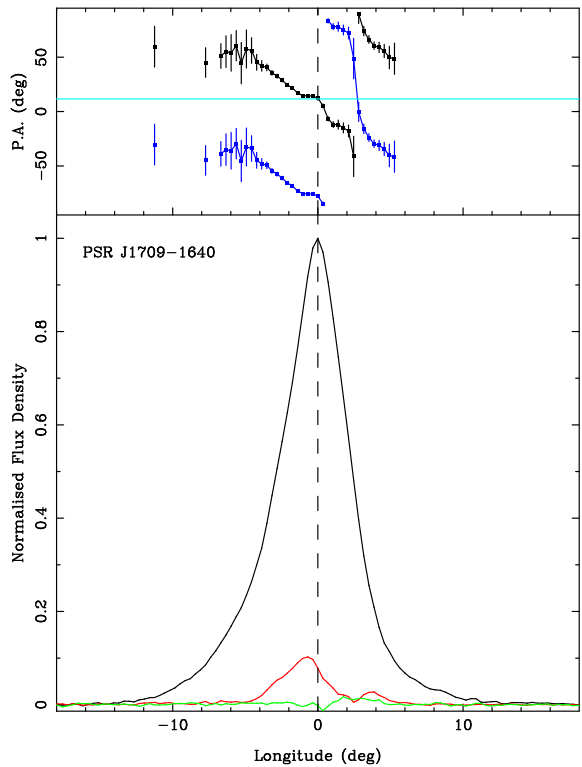}
\caption{Polarization profile of B1706--16 from Johnston I showing its highly 
depolarized single core-component profile at 1369 MHz.  The roughly linear, 
negative-going PPA (black curve) plotted in the upper panel connects smoothly 
with the  PPA -- 90\degr\ (blue) curve on the trailing edge of the profile, coinciding 
with the minimum in the linear polarization (red curve) in the lower panel where 
the total power (black) and circular polarization (green) are also plotted.  The PPAs 
are referred to infinite frequency, and the $PA_V$ value is shown as a horizontal 
(cyan) line.  See text.}
\label{fig2}
\end{center}
\end{figure}

In large part this failure stemmed from the complexity of the PPA traverses in some 
pulsars with bright core components.  A number exhibit PPA traverses that are readily 
fitted by the expected rotating-vector (RVM) model \citep{1969ApL.....3..225R}, 
while others have depolarized, OPM-active components with distorted PPA traverses 
that have proven  difficult to explain and interpret.  Recently, several detailed analyses 
of particular bright pulsars with prominent core emission have helped us to understand 
these effects more fully [\cite{2007MNRAS.379..932M}; \cite{2013MNRAS.435.1984S}], 
and we have found that intensity-dependent aberration/retardation 
\citep[A/R, see][]{1991ApJ...370..643B}---an effect first seen in the Vela pulsar 
\citep{1983ApJ...265..372K}---is also a major source of PPA-traverse distortion in cores.  

Figure~\ref{fig2} shows a clear case of this sort of depolarization and distortion of an 
otherwise nearly linear PPA traverse.  The emission in the trailing part of the component 
reflects one OPM and that in the earlier part the other, apparently because more intense 
fractions of this later ``parent'' emission appear earlier due to the intensity-dependent 
A/R and some of it is seemingly converted to the other OPM.  

\begin{figure}
\begin{center}
\includegraphics[width=80mm, angle=-90]{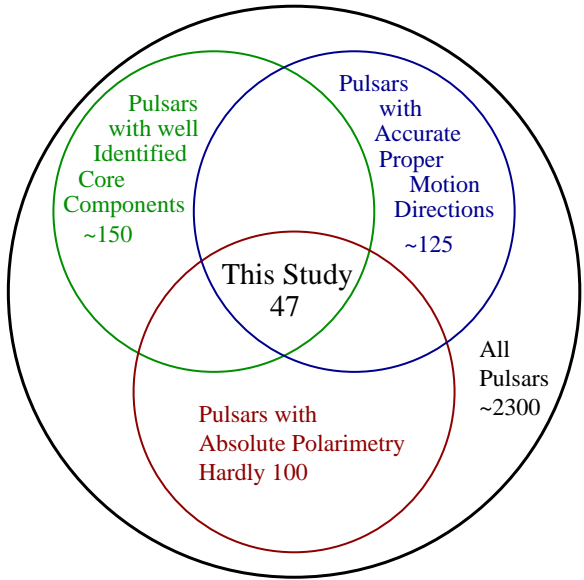}
\caption{Venn diagram showing the particular population of 47 pulsars having 
accurate determinations of three different types that is the foundation of this analysis.}
\label{fig3}
\end{center}
\end{figure}

\section{Three Overlapping Analyses}  The analysis of this paper is based on 
three different types of pulsar investigation: a) reliable identification of pulsars with 
central core-beam emission; b) well determined fiducial PPA measurements; and 
c) accurate proper motion (PM) directions.  

Core emission was first distinguished from the conal type in ET I and populations of 
pulsars with core emission components identified in ET IV, VI \& IX.  Emission geometry 
analyses for more than 50 core single ({\bf S$_t$}), some 50+ core-cone triple ({\bf T}) 
as well as a few core/double-cone ({\bf M}) stars are given in the tables of ET VIb.  This 
overall core population is key to our analysis below because almost all of these 
identifications have proven to be accurate, many through further detailed single pulse 
studies.  Moreover, core components usually appear to fall close to the magnetic axis 
longitude in pulsar profiles, so provide a useful indicator in this regard.  

Fiducial polarization angle measurements $PA_0$ are here used as proxies for the 
orientation of a pulsar's rotation axis on the sky.  First, they require absolute PPA 
calibration---that is, referenced to an origin measured counter-clockwise from North 
on the sky---and second, they must be referred to infinite frequency by accurately 
unwrapping the Faraday rotation of the PPAs.  Then the ``fiducial'' magnetic axis 
longitude is estimated by PPA-fitting or profile-analysis.  Most of the published pulsar 
polarimetry lacks absolute calibration.  The Effelsberg instrument pioneered this work 
[\cite{1979A&A....73...46M}; \cite{1981A&AS...46..421M}; \cite{1991A&A...241...87X}], 
but important recent efforts have been carried out at Parkes (Johnston I; 
\cite{2006MNRAS.365..353K}; \cite{2007MNRAS.381.1625J}, hereafter Johnston II]
 as well as Arecibo (Paper I) and the GBT [\cite{2012MNRAS.423.2736N}; 
\cite{preprint}]---bringing the total to about 100 objects.   

Accurate proper-motion directions $PA_V$, measured CCW from North, are known 
for just over 100 normal and some 25 millisecond pulsars.  Earlier, pulsar PMs were 
measured reliably only with interferometry, but over the last decade special pulsar-timing 
methods have also produced useful measurements (Hobbs \etal\ 2004, 2005).  

\begin{table}
\begin{center}
\caption{Alignment Angles for Pulsars with Core Components.\label{tbl-1}}
\smallskip
\footnotesize
\begin{tabular}{llllll}
\tableline\tableline
Pulsar & $PA_V$ &Ref.& $PA_0$  &Ref.& $\Psi$  \\
 & (deg) && (deg) && (deg) \\ 
\tableline
\\
B0011+47 & +136(3)&1& 43(7) &\ref{tbl-A4}& --87(8) \\ 
B0136+57 & --131(0)&2 & 43(3) &\ref{tbl-A4}& 6(3) \\
B0329+54 & 119(1)&3&  20(4)  &\ref{tbl-A2}&   99(4)  \\
B0355+54 &  48(1)&4&  --41(4) &\ref{tbl-A2}&   89(5)  \\ \smallskip
B0450+55 & 108(0)&2 & --23(16) &\ref{tbl-A2}& --94(16) \\ 
B0450--18 & 40(5)&2 & 47(3) &\ref{tbl-A3}& --7(6) \\
B0540+23 &  58(19)&5& --85(3) &\ref{tbl-A3}& --37(19) \\
B0628--28 & 294(2)&1 &  26(2) &\ref{tbl-A1}& 88(3) \\
B0736--40 & 227(5)$^{\it a}$&1& --44(5) &\ref{tbl-A2}& 91(7) \\ \smallskip
B0823+26 & 146(1)&6& 48(3)  &\ref{tbl-A2}&  98(3)  \\
B0833--45 & 301(0)&7 & 37(1) &\ref{tbl-A1}& 84(1) \\
B0835--41 & 187(6)&1 &  --84(5) &\ref{tbl-A3}& 91(8) \\
B0919+06 & 12(0)&8  & --77(10) &\ref{tbl-A1}& 89(10) \\
B1237+25 & 295(0)&3 & --28(4) &\ref{tbl-A1}& --37(4) \\ \smallskip
B1322+83 &--76(13) &5& 26(3) &\ref{tbl-A4}& 78(15) \\
B1426--66 & 236(5)&9 & --29(1) &\ref{tbl-A1}& 84(5)  \\
B1449--64 & 217(2)&9 & --57(0) &\ref{tbl-A1}& 94(2) \\
B1451--68 &  253(0)&10 & --22(4) &\ref{tbl-A1}& 95(4) \\
B1508+55 & --130(0)&5&   5(4)   &\ref{tbl-A2}& 45(4) \\ \smallskip
B1541+09 &  --111(0)&2& --32(15) &\ref{tbl-A5}&  --79(17) \\
B1600--49 & 268(6)&9&  0(5) &\ref{tbl-A3}& 88(8) \\
B1642--03 & 353(3)&1 & 89(8) &\ref{tbl-A1}& 84(8) \\
B1706--16 & 192(4)&11 & --75(2) &\ref{tbl-A1}&  87(4) \\
B1706--44 & 160(10)&12 & 71(10) &\ref{tbl-A1}& 89(14) \\ \smallskip
B1732--07 & --5(3)&1&  --22(1) &\ref{tbl-A4}& 17(4) \\
B1737+13 & 228(6)&1 & --46(4) &\ref{tbl-A1}& 94(7) \\
B1818--04 & --22(17)&13& 55(3) &\ref{tbl-A3}& --77(17) \\
B1821--19 &--173(17)&14 &53(2)&\ref{tbl-A4}& --46(17) \\
B1826--17 &+172(9)&14 &11(1)&\ref{tbl-A4}& --19(9) \\ \smallskip
B1842+14 & 36(8)&11 & --52(2) &\ref{tbl-A1}& 88(9) \\
B1848+13 & 237(16)&13& --45(3) &\ref{tbl-A3}& --78(16) \\
B1857--26 & 203(0)&5 & --69(10) &\ref{tbl-A1}& 92(10) \\
B1911--04 & 166(5)&11 & --99(8)  &\ref{tbl-A1}& 85(9) \\
B1913+10 & 174(15)&13&  85(3) &\ref{tbl-A3}& 89(15) \\ \smallskip
B1929+10 & 65(0)&8 & --11(0) &\ref{tbl-A1}& 77(0) \\
B1933+16 & 176(1)&11 & --87(1) &\ref{tbl-A1}& 89(1) \\
B1935+25 & 220(9)&13&  --56(8) &\ref{tbl-A3}& 96(12) \\
B1946+35 & --93(3)$^{\it b}$&13& --78(15) &\ref{tbl-A5}& --16(15) \\
B2045--16 & 92(2)&15&  --13(5) &\ref{tbl-A3}& --75(6) \\ \smallskip
B2053+36 & 157(1)&5& --77(11) &\ref{tbl-A5}& 54(11) \\
B2110+27 & --157(2)&5&--64(5) &\ref{tbl-A5}& 87(5) \\
B2111+46 & --20(44) &13 & 86(0) &\ref{tbl-A4}& 66(44) \\
B2217+47 &--158(10)&6 &  65(8) &\ref{tbl-A4}& --4411)   \\
B2224+65 & +52(1) &5& --48(5) &\ref{tbl-A4}& --80(5) \\ \smallskip
B2255+58 & +106(12)&14 & 24(3) &\ref{tbl-A4}& 82(12) \\ 
B2310+42 & +76(0) &2 & 18(10) &\ref{tbl-A4}& 58(10) \\
B2327--20 &  86(2)&1& 21(10) &\ref{tbl-A3}& 65(10) \\ 
\tableline
\end{tabular}
\end{center}
\scriptsize
$^a$Here we use \cite{2003AJ....126.3090B} $PA_V$ value rather than the analysis in 
	Johnston \etal\ II, correcting the error in Paper I.
{$^b$}The origin of the interferometric proper motion value in \cite{2005MNRAS.360..974H}  
is unclear, so we have used the timing value in Hobbs \etal\ (2004).

Proper-motion references: (1) \cite{2003AJ....126.3090B}, (2) \cite{2009ApJ...698..250C},
(3) \cite{2002ApJ...571..906B}, (4) \cite{2005ApJ...630L..61C}, (5) \cite{1993MNRAS.261..113H}, 
(6)\cite{1982MNRAS.201..503L}, (7) \cite{2003ApJ...596.1137D}, (8) \cite{2001ApJ...550..287C}, 
(9) \cite{1990Natur.343..240B}, (10) \cite{1990MNRAS.247..322B}, 
(11) Johnston \etal\ I updates of Hobbs \etal\ (2004), (12) \cite{2004ApJ...601..479N}, 
(13) \cite{2004MNRAS.353.1311H}, (14) \cite{2005MNRAS.362.1189Z}, 
(15) \cite{1997MNRAS.286...81F}

\normalsize
\end{table}

In what follows then, we assemble this three-fold information as indicated in 
Figure~\ref{fig3} on the nearly 50 pulsars having reliable PM directions, absolute 
fiducial PPA determinations and well identified core-emission components. 
Table~\ref{tbl-1} summarizes these proper-motion directions $PA_V$, proxy 
rotation-axis orientations $PA_0$ as well as their differences $\Psi$.  The $PA_V$ 
references show the origins of these values and in several cases their correction.  
The $PA_0$ references trace their origins from Johnston I, Paper I, Johnston II, Force \etal, 
and the present study (Tables \ref{tbl-A1}-\ref{tbl-A5}, respectively).  Notes to these Appendix 
tables explain needed revisions or problems.  Most classifications remain accurate from 
ET IV and ET VI, with only a few having been classified anew or reclassified as a result of 
new information.  Similarly, most $PA_0$ values appeared accurate as determined in the 
above sources, and where corrections or different interpretations are made, these are 
described in the table notes.  

\begin{figure}
\includegraphics[width=76mm]{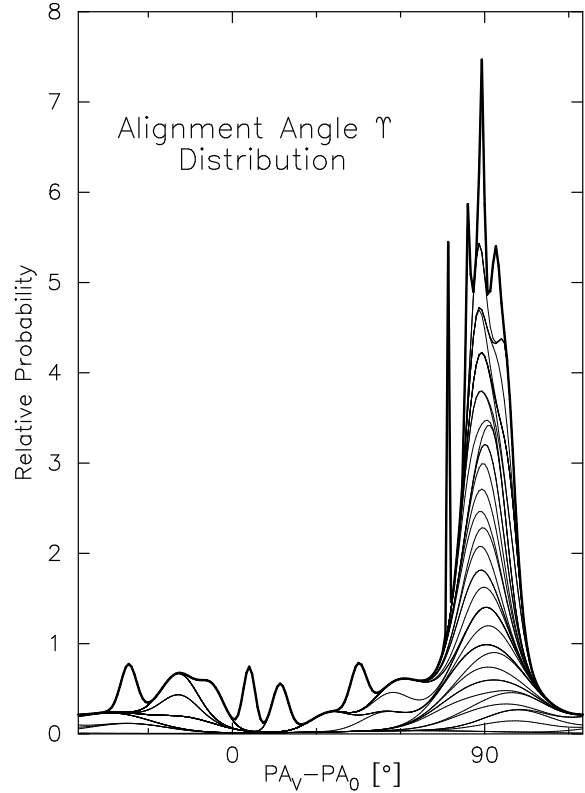}
\caption{Distribution showing the core-emission alignment angles $\Psi$ of the 47 pulsars 
in Tables~\ref{tbl-A1} to \ref{tbl-A5}.  Here the alignment of each pulsar is represented by  
a 180\degr-von Mises function with peak $\Psi$ and standard deviation corresponding to its 
error value.  Each function has equal area such that the total area of the cumulative distribution 
corresponds to the area of the region below unity.}
\label{fig4}
\end{figure}

\section{Discussion} 
The overall distribution of the computed alignment angles $\Psi$ [= $PA_V - PA_0$] for core 
emission is shown in Figure~\ref{fig4}.  Each value is represented by a von Mises function 
of equal area with its standard deviation corresponding to its error.  All 47 pulsars currently 
having the three-fold analysis---that is, accurate proper motions together with fiducial PPAs of 
relatively well identified core emission---appear in Table~\ref{tbl-1}, the Appendix tables and 
the above figure.  

Clearly, the analysis shows that core emission exhibits a strong orthogonal alignment on the 
plane of the sky.  Most of the $\Psi$ values fall near 90\degr\ and often accurately so.  Indeed, 
interpreting Fig.~\ref{fig4} probabilistically, half the weight falls within 90$\pm$10\degr, 
2/3 within $\pm$20\degr and 3/4 within $\pm$30\degr; only 14\% falls within 0$\pm$30/degr. 
We see here that the ``fiducial'' PPA $PA_0$ (in several cases computed to track the ``parent'' 
core OPM) is usually orthogonal to the proper-motion direction $PA_V$.  

The ``fiducial'' instant furthermore implies that the projected direction of the emitting 
magnetic field is in turn $\parallel$ to the rotation axis \Omga\ on the sky (see Fig.~\ref{fig1}).  
None of this, however, fixes how the electric vector (defining the linear PPA direction) 
of the core emission is oriented with respect to the projected magnetic field direction.  
For this we must refer to X-ray images of Vela and certain other pulsars (\eg, Helfand 
\etal\ 2001; Ng \& Romani 2004) from which the rotation-axis orientation on the sky can 
be compared with the radio ``fiducial'' polarization direction of the core emission---and the 
result that they are again orthogonal.  

These circumstances then support several conclusions---
\begin{itemize}
\item Core emission tends to be polarized perpendicularly to the projected magnetic 
field direction and thus propagates as the extraordinary (X) physical mode.  In a few 
cases we must distinguish the (later or probably lower altitude)  ``parent'' core emission.
\item Most pulsar proper motions fall closely $\parallel$ to the magnetic axis direction 
\Omga.  The narrowness of this distribution around 90\degr\ is first surprising given the 
mechanisms that would degrade such alignments, and second a $\parallel$ alignment 
is not what would be expected if binary disruption is a major contributor to pulsar proper 
motions. 
\item Natal supernovae ``kicks'' would then seem to be the dominant mechanism behind 
pulsar proper motions, and these ``kicks'' are primarily directed $\parallel$ to a pulsar's
rotation axis.   
\item The orderly orientations of core polarization provide further demonstration of the 
the distinct character of core emission by contrast to conal radiation or other types not 
yet identified.  
\item That core radiation both is the primary (central, low altitude) radio-frequency emission 
we are able to detect and tends strongly to propagate as the extraordinary (X) mode 
provides major insights into the operation of the pulsar ``engine''.  It must be tertiary to 
the high energy primary particle acceleration just above the polar cap.  It reflects heavy 
processing by the highly magnetized dense plasma within the polar flux tube above the 
polar cap in which electromagnetic waves cannot propagate.  Therefore, it must originate 
at a somewhat higher altitude at which plasma wave coupling to the X mode first becomes 
possible.  
\end{itemize}

These conclusions are important not only for understanding the origins of pulsar emission 
but also for supernova theory.  Spin-aligned supernova ``kicks'' imply either rotation averaging 
of hydrodynamic ``kicks'' on timescales much longer that the natal spin period or a magnetic 
field driven mechanism.  

Similarly, the orderly alignments have important implications for understanding the 
detailed characteristics of the emission from individual pulsars.  As we saw in our 
study of pulsar B0329+54 (Mitra \etal\ 2007), our ability to identify the X and O modes 
of the emission facilitates a much more physical interpretation.

Establishing that the primary core emission is X mode further enables us to distinguish 
the magnetic orientation of its two OPMs everywhere within a pulsar's profile and to pursue 
questions about the overall emission more physically.  Or said differently, the identification 
of core-associated X-mode emission in the profile centers of triple or five-component 
profiles permits us to identify the X/O character of the two conal-associated OPMs. 

This all said, our largely average-polarimetry based analysis above remains the result 
of a rough tool.  Our facilitating insights have come from the few available pulse-sequence 
polarization studies, and very much remains to be learned from such investigations of 
other pulsars.  As regards core emission, its mechanisms and dynamics will be revealed 
through broad band polarimetric studies of its depolarization.  

While we have been fairly successful in identifying core features on the basis of their 
geometry, spectral behavior and circular polarization, their linear polarization properties 
are highly variable.  A few are highly polarized, but many are not, some are unimodal in 
form and other show bifurcation or a leading ``pedestal'' feature.  In view of these varying 
characteristics, again it is surprising that the distribution of alignments is as strongly 
orthogonal as we have seen in Fig.~\ref{fig4}.  In particular, the ``parent'' core emission 
is not always dominant or is conflated with other profile features so is not easy to identify.  
Thus further study may well show that some of the 0\degr\ alignments represent O-mode 
emission rather than the ``parent'' X-mode radiation.

Again, we emphasize that these alignments are on the sky, as we have no direct means 
of considering the radial components of pulsar velocities.  Given, however, the surprisingly 
compact distribution of transverse alignments, it is hard to imagine that the radial alignments 
could have a very different distribution.  The effect of the unknown radial velocity components 
can be modeled statistically as Noutsos \etal\ (2012) have done, and their overall conclusion 
to the above effect suggests that a detailed statistical modeling is unnecessary to support 
the results here.  

It is interesting to look closely at those pulsars showing $\parallel$ or oblique alignments.  
Regarding the latter, some undoubtedly are cases like B1237+25 and B1508+55 where their 
positions on the sky are such that our lack of the radial velocity component is crucial---and 
the $\Psi$ values thus misleading.  Several other pulsars undoubtedly remain misclassified, 
for instance because core ({\bf S$_t$}) and conal single ({\bf S$_d$}) profiles can sometimes 
be difficult to distinguish.  

The overall implication of our interpretation is that the fundamental curvature emission drives 
longitudinal plasma oscillations in the dense inner plasma of the polar flux tube \citep[\eg,][]{2002nsps.conf..230L}.  These plasma oscillations then first couple to X-mode radiation which often then seems 
to be converted into the O mode at higher altitude in an intensity-dependent manner.  The 
plasma oscillations may well be responsible for the ``bunching'' needed for pulsars to radiate 
mainly at radio frequencies, and the consistent characteristic dimensions of core components 
probably follows from the non-refractiveness of the X mode.

\noindent
{\bf Acknowledgments}: The author thanks the anonymous referee and also Alice Harding, 
Aris Noutsos, Dipanjan Mitra and Paul Demorest for helpful suggestions on the analysis and 
manuscript.  Portions of this work were supported by US NSF grants 08-07691 and 09-68296.  
Arecibo Observatory is operated by SRI International under a cooperative agreement with the 
NSF, and in alliance with Ana G. M\'endez-Universidad Metropolitana, and the Universities 
Space Research Association.  This research has made use of NASA's Astrophysics Data 
System.

\bibliographystyle{apj}
\bibliography{coreXmode}

\label{lastpage}

\appendix
\setcounter{figure}{0} 
\renewcommand{\thefigure}{A\arabic{figure}}
\renewcommand{\thetable}{A\arabic{table}}
\setcounter{table}{0} 
\renewcommand{\thefootnote}{A\arabic{footnote}}
\setcounter{footnote}{0} 
\section*{APPENDIX: Results for Individual Pulsars}
\label{sec:results}
%\twocolumn

\begin{table}
\begin{center}
\caption{Fiducial Polarization Angles for Stars Studied by Johnson I.\label{tbl-A1}}
\smallskip
\footnotesize
\begin{tabular}{llcllc}
\tableline\tableline
Pulsar &Cl & $P$ &log($\tau$)& $PA_0$  & Method  \\
 && (s) & (yrs)& (deg) &    \\
\tableline
B0628--28$^{\it a}$ &{S$_t$}& 1.244&6.443&  26(2) & RVM fit \\
B0833--45 &{S$_t$}& 0.0893&4.053& 37(1) & RVM fit \\
B0919+06 &{T} & 0.4306&5.696& --77(10) &  PPA geom.$^{\it b}$ \\
B1237+25 &{M} & 1.3824&7.358& --28(4) &  traverse \\
B1426--66 &{T?} & 0.7854&6.652& --29(1) &  RVM fit \\ \smallskip
B1449--64 &{S$_t$} & 0.1795&6.017& --57(0) & RVM fit \\
B1451--68 &{M}& 0.2634&7.628& --22(4) & RVM fit \\
B1642--03 &{S$_t$} & 0.3877&6.538& {\bf 89(8)}$^{\it c}$ & see text \\
B1706--16 &{T} &0.6531&6.215& {\bf --75(2)}$^{\it c}$ & see text \\
B1706--44 &{S$_t$} & 0.1025&4.243& 71(10) & Paper I  \\ \smallskip
B1737+13 &{M} & 0.8031&6.943& --46(4) & RVM fit \\
B1842+14 &{T} & 0.3755&6.502& --52(2) & RVM fit \\
B1857--26 &{M} & 0.6122&7.676& --69(10) & Paper I \\
B1911--04 &{S$_t$} & 0.8259&6.508& --99(8)  & Paper I  \\
B1929+10 &{T} & 0.2265&6.491& --11(0) &  RVM fit \\ \smallskip
B1933+16 &{T} & 0.3587&5.976& {\bf --87(1)}$^{\it d}$ & see text \\
\tableline
\end{tabular}
\end{center}
\scriptsize
$^a$ET VI classified this star as a conal (S$_d$) profile, but newer information 
now tilts toward its being a S$_t$ star.  Like B0823+26, it shows neither profile 
bifurcation nor conal outriders.  
$^b$ See \cite{2006MNRAS.370..673R}
$^c$The pulsar's latest core emission is polarized along a ``track'' that is $\perp$ 
to the PM direction at the profile peak.  See text.

$^d$Here the underlying SPM PPA sweep is clearly seen only prior to about --7\degr\ 
longitude and in the tailing core region near +5-7\degr---and connecting the two 
shows the main regions of PPM power under the main peak and in the --6 to --3\degr\ 
interval.  Connecting the former gives and intercept of about --87\degr\ at the primary 
profile peak.  See the text.

\normalsize
\end{table}

\begin{table}
\begin{center}
\caption{Fiducial Polarization Angles for Stars Studied in Paper I.\label{tbl-A2}}
\smallskip
\footnotesize
\begin{tabular}{llcllc}
\tableline\tableline
Pulsar &Cl & $P$ &log($\tau$)& $PA_0$  & Method  \\
 && (s) & (yrs)& (deg) &  \\
\tableline
B0329+54 &{T/M} & 0.7145&6.743&  20(4)  & PPA geom.  \\
B0355+54 &{S$_t$} & 0.1564&5.751&  --41(4) & PPA geom. \\
B0450+55 &{T} & 0.3407&6.358& --23(16)$^{\it a}$ & PPA geom. \\
B0736--40 &{T} & 0.3749&6.566& --44(5) & PPA geom. \\
B0823+26 &{S$_t$} & 0.5307&6.692& 48(3)  & PPA geom. \\
B1508+55 &{T} & 0.7397&6.369&   5(4)   & PPA geom. \\
\tableline
\end{tabular}
\end{center}
\scriptsize
$^a$Both Xilouris \etal\ (1991) and recently Noutsos \etal\ (2012) provide absolute 
polarimetry and in neither is the central PPA rotation resolved; however, the 327-MHz 
polarimetry in ET IX as well as that of \cite{1998MNRAS.301..235G} and 
\cite{1988SvA....32..177S} show that the 
PPA rotates negatively though about 140\degr.  Moreover, strong A/R effects in this 
pulsar were documented in ET IX from which it is clear that the fiducial longitude lags 
the central core by some 13\degr.  Therefore a reasonable estimate of the PPA at the 
fiducial longitude $PA_0$ is about --23\degr.

\normalsize
\end{table}

\subsection*{Johnston I and Paper I Values}
Of the 25 pulsars analyzed by Johnston \etal\ (2006, Johnston I) and then 
reanalyzed by Rankin (2007, Paper I), 16 appear in Table~\ref{tbl-A1} by 
virtue (with one exception) of their earlier classification in ET IV and/or 
ET VIb as having a core emission component.  Apart from pulsar B1237+25, 
they all exhibit alignments $\Psi=PA_V-PA_0$ of close to 90\degr.  The 
alignment values for B1642--03, B1706--16 and B1933+16 have 
been reinterpreted here as we have discussed above.  They are given in 
boldface and differ from the values in Paper I in a similar manner as 
explained below.  

These and several other core-dominated pulsars in Johnston I (\eg, B1426--66 
and B1451--68) share the common properties of distorted PPA traverses and 
highly depolarized profiles.  In Paper I we appealed to observations at other 
(often lower) frequencies to more reliably interpret these PPA traverses.  Now, 
dynamical studies of two similar pulsars [B0329+54 (Mitra \etal\ (2007) and 
B1237+25 (Smith \etal\ (2013)] have helped us understand how this distortion 
and depolarization occurs systematically.  Single mode emission in the trailing 
parts of a core component also appears earlier because of intensity-dependent 
aberration/retardation, and in some cases it seems to undergo conversion to 
the other mode.  A relatively straightforward such case is shown for pulsar 
B1706--16 in Figure~\ref{fig2}, where overall the emission in the two OPMs 
follows a negative-going, nearly linear PPA traverse.  Here and in other cases, 
this later core emission (not that of any trailing conal outrider) is relatively 
undistorted by the intensity-dependent A/R and consequent depolarization 
through OPM mixing, so we use this later OPM to estimate the ``fiducial'' PPA.  
Clearly core emission entails a variety of core properties wherein for Vela and 
most other pulsars in Table~\ref{tbl-A1} the dominant (or later ``parent'') emission 
is $\perp$ to the PM $PA_v$ direction and thus to {\bf B}; whereas in other cases 
complex depolarization occurs and/or the secondary, apparently converted OPM 
becomes dominant at the fiducial longitude.  

Paper I also studied pulsars using the absolute polarimetry from  Morris \etal\ 
(1979, 1981), Xilouris \etal\ (1991) and Karastergiou \& Johnston (2006).  Of the 
21 such stars in Paper I, 6 appear in Table~\ref{tbl-A2} below again by virtue of 
their classification among one of the three profile groups having core components 
in ET IV or VIb.  For these prominent and well studied pulsars,  the PPA traverse 
geometry is usually well understood from multiple sources.  Most of these $PA_0$ 
values are identical to those in Paper I; however, the $PA_0$ values for B0450+55 
has been reinterpreted on the basis of new information.  

\begin{table}
\begin{center}
\caption{Fiducial Polarization Angles for Stars Studied by Johnson II.\label{tbl-A3}}
\smallskip
\footnotesize
\begin{tabular}{llcllc}
\tableline\tableline
Pulsar &Cl & $P$ &log($\tau$)& $PA_0$  & Method  \\
 && (s) & (yrs)& (deg) &  \\
\tableline
B0450--18$^{\it a}$ &{T} & 0.5489&6.179& 47(3) & RVM fit \\
B0540+23$^{\it b}$ &{S$_t$} & 0.2460&5.403& --85(3)$^{\it c}$ & RVM fit \\
B0835--41 &{S$_t$} & 0.7516&6.526&  --84(5) & RVM fit \\
B1600--49 &{T?} & 0.3274&6.707&  0(5)$^{\it d}$ & $V$ zero \\
B1818--04 &{T} & 0.5981&6.176& 55(3)$^{\it e}$ & PPA geom. \\ \smallskip
B1848+13 &{S$_t$}? & 0.3456&6.565& --45(3) & RVM fit \\
B1913+10 &{S$_t$} & 0.4045&5.623&  85(3) & RVM fit \\
B1935+25 &{S$_t$} & 0.2010&6.695&  --56(8)$^{\it f}$ & PPA geom. \\
B2045--16 &{T} & 1.9616&6.453&  --13(5) & RVM fit \\
B2327--20 &{T}&  1.6436&6.750& 21(10) & RVM fit \\
\tableline
\end{tabular}
\end{center}
\scriptsize
{$^a$} This pulsar needs study at a single pulse level.  Johnston II's 3.1-GHz and
and 691-MHz profiles seem from different stars---and the smooth high frequency 
PPA traverse is deceptive; most lower frequency profiles show multiple $L$ minima 
and 90\degr\ ``jumps''---possibly due to A/R effects. 

{$^b$}Long classified as {\bf S$_t$}, the asymmetry of the PPA inflection 
and lack of outriders cast this classification into serious doubt.  

{$^c$} BCW's fiducial PPA longitude falls 18\degr\ after  the peak.

{$^d$}Fiducial longitude taken at the $V$ zero-crossing point.
	
{$^e$}The high frequency core center provides a better fiducial longitude.

{$^f$}This pulsar's leading component is a highly polarized precursor, and the second 
feature appears to have a core-single profile.  Thus we take the fiducial longitude at 
what seems to be the core peak at 3.1 GHz.  The 691-MHz profile is so depolarized 
in this region that no reliable PPAs can be measured.

\normalsize
\end{table}

\begin{table}
\begin{center}
\caption{Fiducial Polarization Angles from Force \etal\ (2015).\label{tbl-A4}}
\smallskip
\footnotesize
\begin{tabular}{llcllc}
\tableline\tableline
Pulsar &Cl & $P$ &log($\tau$)& $PA_0$  & Method  \\
 && (s) & (yrs)& (deg) &  \\
\tableline
B0011+47 &{T?} & 1.2407&7.542& 43(7) & RVM fit \\
B0136+57$^a$&{S$_t$} & 0.2725&5.605& 43(3) & RVM fit \\
B1322+83 &{S$_t$} & 0.6700&7.272& 26(3) & PPA geom. \\ 
B1732--07 &{T} & 0.4193&6.738&  --22(1)$^{\it b}$ & RVM fit \\
B1821--19 &{S$_t$}?& 0.1893&5.758&53(2)& RVM fit \\
B1826--17 &{T?} & 0.3071&5.943 &11(1)& RVM fit \\
B2111+46 &{T}  & 1.0147&7.352& 86(0) & RVM fit \\
B2217+47 &{S$_t$} & 0.5385&6.490&  65(8) & RVM fit \\
B2224+65 &{S$_t$} & 0.6825&6.049& --48(5) & ET IX,X \\
B2255+58 & {S$_t$} & 0.3682&6.004& 24(3) & RVM fit \\
B2310+42 & {M} & 0.3494&7.693& 18(10) & RVM fit \\
\tableline
\end{tabular}
\end{center}
\scriptsize
$^a$ ET VIb classified this star as having an {\bf S$_t$} profile, but 
\citet[\citeyear{2007A&A...469..607W}]{2006A&A...445..243W} find weak evidence of drifting.  
Neither conal outriders nor low frequency bifurcation has been seen---so the classification 
is uncertain. The PPA rotation in Noutsos \etal\ (2012) seems to have the incorrect sense.
	
{$^b$}Fitted steep central PPA traverse unresolved in Johnston II.
	
\normalsize
\end{table}

\begin{table}
\begin{center}
\caption{Fiducial Polarization Angles from this paper.\label{tbl-A5}}
\smallskip
\footnotesize
\begin{tabular}{llcllc}
\tableline\tableline
Pulsar &Cl & $P$ &log($\tau$)& $PA_0$  & Method  \\
 && (s) & (yrs)& (deg) &  \\
\tableline
B1541+09 &{T} & 0.7484&7.438& --32(15) & PPA geom.  \\
B1946+35$^{\it a}$ &{S$_t$}& 0.7173&6.207& --78(15) & PPA geom. \\
B2053+36 &{S$_t$} & 0.2215&6.978& --77(11) & PPA geom. \\
B2110+27 &{S$_t$} & 1.2028&6.861&--64(5) & PPA geom. \\
\tableline
\end{tabular}
\end{center}
\scriptsize
{$^a$}Single-pulse polarimetry study is needed to reliably interpret this pulsar's 
complex and depolarized profile in the core region.  
	
\normalsize
\end{table}

\begin{figure*}
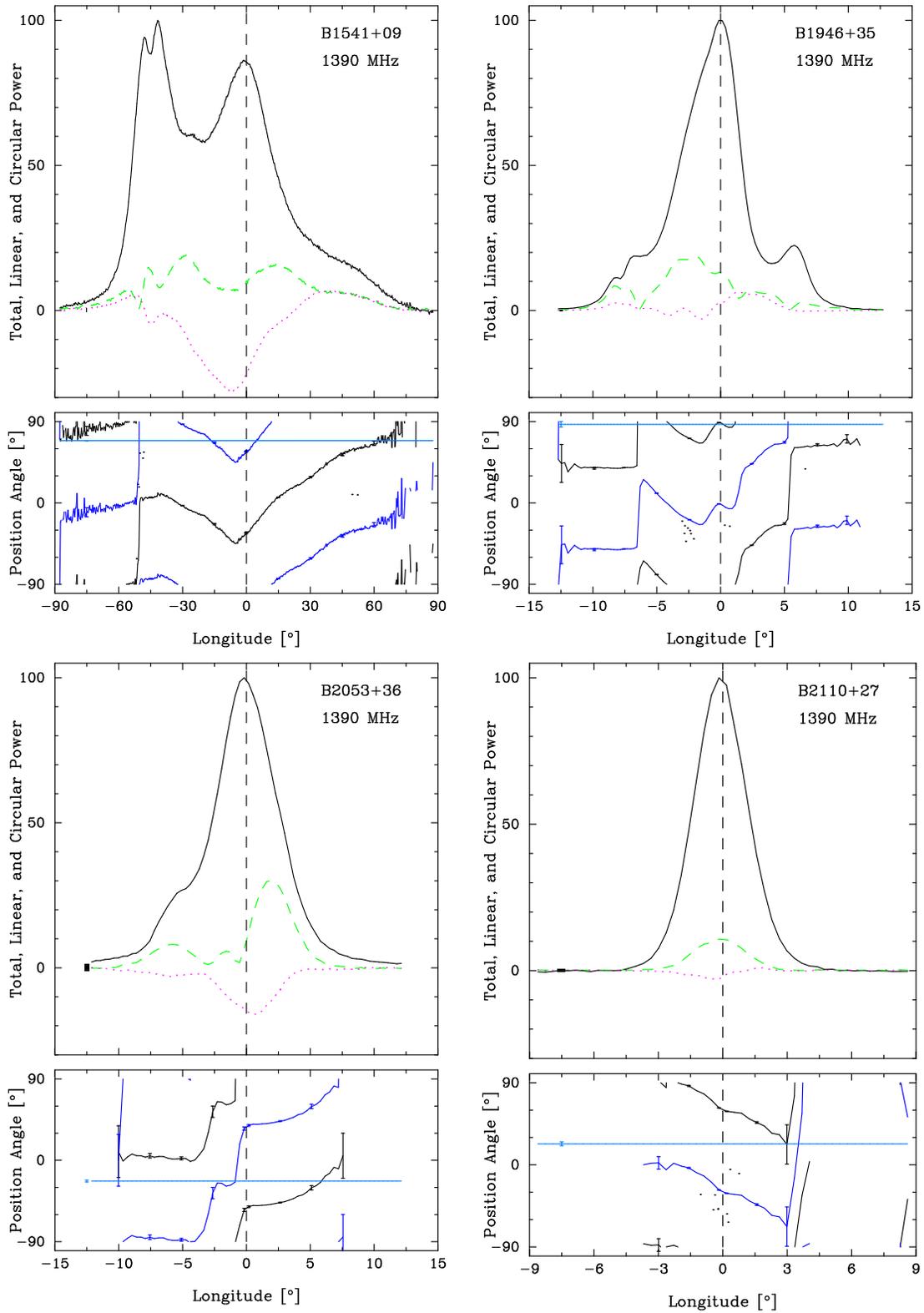

\begin{center}
\begin{tabular}{@{}cc@{}c}
{\mbox{\includegraphics[height=70mm,angle=-90.]{PQPMPA_B1541+09.55637la.ps}}}& \ \
{\mbox{\includegraphics[height=70mm,angle=-90.]{PQPMPA_B1946+35.55632l.ps}}}\\ \smallskip\smallskip
{\mbox{\includegraphics[height=70mm,angle=-90.]{PQPMPA_B2053+36.55632l.ps}}}&\ \
{\mbox{\includegraphics[height=70mm,angle=-90.]{PQPMPA_B2110+27.55632l.ps}}}\\
\end{tabular}
\caption{Polarization profiles for B1541+09, B1946+35, B2053+36 and B2110+27.  
Here the total power (black), total linear polarization (green dashed) and circular 
polarization (magenta dotted) are plotted in the upper panels.  The Faraday derotated 
PPA information (black) is plotted in the lower panels and the PPA--90\degr\ (blue) 
is also shown along with bars indicating the errors.  Finally, a cyan line shows the 
$PA_V$ value.} 
\label{figA1}
\end{center}
\end{figure*}

\subsection{Johnston II, Force \etal, and AO Values}
A further major source of fiducial $PA_0$ values is Johnston \etal\ (2007, 
Johnston II), and these appear in Table~\ref{tbl-A3}.   Of the 22 pulsars in 
the foregoing paper, 10 appear here, and most were classified as above 
or in ET IX.  Overall this group is less well studied in terms of profile geometry, 
often because polarimetric observations are available only at one or two 
frequencies.  Notes show how the pulsars were interpreted here when the 
resulting $PA_0$ values differ from those in Johnston II.

Force \etal\ (2015) conducted absolute polarimetric measures  with the 
Green Bank Telescope.  Of the 33 Force \etal\ stars, 11 with clear or 
probable core components  appear in Table~\ref{tbl-A4}.  Again, this 
group has been less well studied in terms of profile geometry, often 
because their weakness makes observations over a broad band difficult.   

Finally, we report four values from Arecibo measurements in this paper.  
These used four Mock Spectrometers (www.naic.edu/$\sim$astro/mock.shtml) 
sampling bands of 86 MHz centered at 1270, 1420, 1520 and 1620 MHz 
with milliperiod sampling.  The resulting polarized profiles were derotated 
to infinite frequency are shown in Fig.~\ref{figA1}.  All reference nominal 
21-cm observations (by averaging the bottom three bands), and they entail 
straightforward interpretations in that their fiducial longitudes fall very close 
to the central component peaks.  Their $PA_0$ values are given in 
Table~\ref{tbl-A5}.

\end{document}